\documentclass{article}

     \usepackage[final]{tackling_climate_workshop_style}

\usepackage{tackling_climate_workshop_style}
\usepackage{graphicx}
\usepackage[utf8]{inputenc} 
\usepackage[T1]{fontenc}    
\usepackage{hyperref}       
\usepackage{url}            
\usepackage{booktabs}       
\usepackage{amsfonts}       
\usepackage{nicefrac}       
\usepackage{microtype}      
\usepackage{enumitem}
\usepackage[font=small,skip=2pt]{caption} 
\usepackage{float}

\usepackage{amsmath}

\usepackage{tcolorbox}
\tcbuselibrary{listingsutf8}
\usepackage{listings}
\usepackage{xcolor}
\usepackage{caption}

\title{\textit{} Reflexive Evidence-Based Multimodal Learning for Clean Energy Transitions: Causal Insights on Cooking Fuel Access, Urbanization, and Carbon Emissions}

\author{
	Shan Shan\\
	Zhejiang University\\
	\texttt{shanshanfy@gmail.com} \\
}

\begin{document}

\maketitle

\begin{abstract}
Achieving Sustainable Development Goal 7 (Affordable and Clean Energy) requires not only technological innovation but also a deeper understanding of the socio-economic factors that influence energy access and carbon emissions. Despite growing attention to these drivers, key questions remain, particularly regarding how to quantify socio-economic impacts, how these impacts interact across domains such as policy, technology, and infrastructure, and how feedback processes shape energy systems.
To address these gaps, this study introduces ClimateAgents, an AI-based framework that combines large language models with domain-specialized agents to support hypothesis generation and scenario exploration. Leveraging 20 years of socio-economic and emissions data from 265  economies, countries and regions, and 98 indicators drawn from the World Bank database, the framework applies a machine learning based causal inference approach to identify key determinants of carbon emissions in an evidence-based, data-driven manner. The analysis highlights three primary drivers: (1) access to clean cooking fuels in rural areas, (2) access to clean cooking fuels in urban areas, and (3) the percentage of  population living in urban areas. These findings underscore the critical role of clean cooking technologies and urbanization patterns in shaping emission outcomes. In line with growing calls for evidence-based AI policy, ClimateAgents offers a modular and reflexive learning system that supports the generation of credible and actionable insights for policy. By integrating heterogeneous data modalities, including structured indicators, policy documents, and semantic reasoning, the framework contributes to adaptive policymaking infrastructures that can evolve with complex socio-technical challenges. This approach aims to support a shift from siloed modeling to reflexive, modular systems designed for dynamic, context-aware climate action.

\end{abstract}

Strengthening science-informed, evidence-based approaches to AI policy is increasingly essential\cite{bommasani2025advancing}.In domains such as clean energy and climate change, the question is how to integrate the strengths of narrow AI components into a unified, context-aware agent, one able to process multimodal data, reason causally, interact socially, and adapt continuously to evolving policy challenges. Recent advances in artificial intelligence (AI), particularly large language models (LLMs), present new opportunities for sustainability research. LLMs are being applied in environmental science, economics, and climate modeling \cite{hadi2024large, binz2025advancement, rillig2023risks, raiaan2024review, MOTLAGHZADEH2023161060, muller2014application, shan2024correlation}, with transformer-based architectures enabling reasoning, inference, and policy analysis \cite{zhou2022large, jin2021mining}. They also support hypothesis generation, in-context retrieval, and multimodal data synthesis \cite{agarwal2024many, luo2024in}. However, sustainability applications remain limited, often constrained to text summarization or keyword-based mapping \cite{zhang2024comprehensive, firoozeh2020keyword, yang2024harnessing, wolf2023fundamental, villalobos2022will}. Current LLMs lack integration with structured data, causal inference, and adaptation to new socio-environmental indicators \cite{bail2024can}. A key barrier is the absence of causal reasoning datasets: while general-purpose corpora from Google \cite{kwiatkowski2019natural}, Bing \cite{nguyen2016ms}, and user interactions (e.g., ShareGPT, WildChat) \cite{zhao2024wildchat} exist, few support causal inquiry in LLM prompts \cite{ouyang2022training, jin2024largelang}. Critically, no dataset addresses socio-climate-related causal questions, leaving a major gap for advanced applications \cite{ceraolo2024}.


To ensure reliable and standardized evidence-based analysis, this research adopts the World Bank Development Indicators—a widely recognized, high-quality, and publicly available data framework. This data-driven foundation enhances the credibility and precision of the study’s outputs.
By applying causal inference techniques with machine learning algorithms, the analysis moves beyond simple correlation to uncover deeper, more robust relationships. This enables more grounded, interpretable reasoning for policy-making. Furthermore, the integration of large language models (LLMs) supports evidence-based analysis by generating outputs that aim to be \textit{credible and actionable}, as their interpretability can facilitate context-aware and informed decision-making \cite{madaan2023self, wiegreffe2021teach, bommasani2021opportunities}.

The orchestration of these three modules reflects a system-level design philosophy rooted in modularity, specialization, and agent-based coordination. To address these challenges and extend the utility of large language models (LLMs), this work proposes a multi-agent architecture grounded in Minsky’s philosophy of modular, emergent intelligence. Rather than treating LLMs as monolithic tools, the proposed system distributes reasoning and task execution across a set of interacting agents, each specialized for distinct  functions.
The resulting framework—ClimateAgents—is a reflexive, causal modeling system powered by GPT  family models \cite{achiam2023gpt} and accessed via the OpenAI Application Programming Interface \cite{openai_api}. It moves beyond static prompt-response paradigms, enabling adaptive reasoning within complex socio-environmental systems. Central to this architecture is the concept of \emph{Reflexive} machine learning, defined here as a process through which agents iteratively adjust their prompts, inference strategies, or actions in response to environmental feedback and task complexity, thereby supporting context-aware and adaptive decision-making.

The contributions of this work are as follows:  
(i) introduction of a reflexive multiagent architecture for causal analysis and policy simulation in socio-environmental contexts;  
(ii) integration of multimodal data with LLM-driven agents to complement statistical models through simulation, reasoning, and hypothesis generation; and  
(iii) proposal of Reflexive Machine Learning as a natural-language interface for interpretable modeling of complex systems.  

\section*{Methods}

\subsection*{ClimateAgents}

\begin{figure*}
\centering
\includegraphics[width=0.8\textwidth]{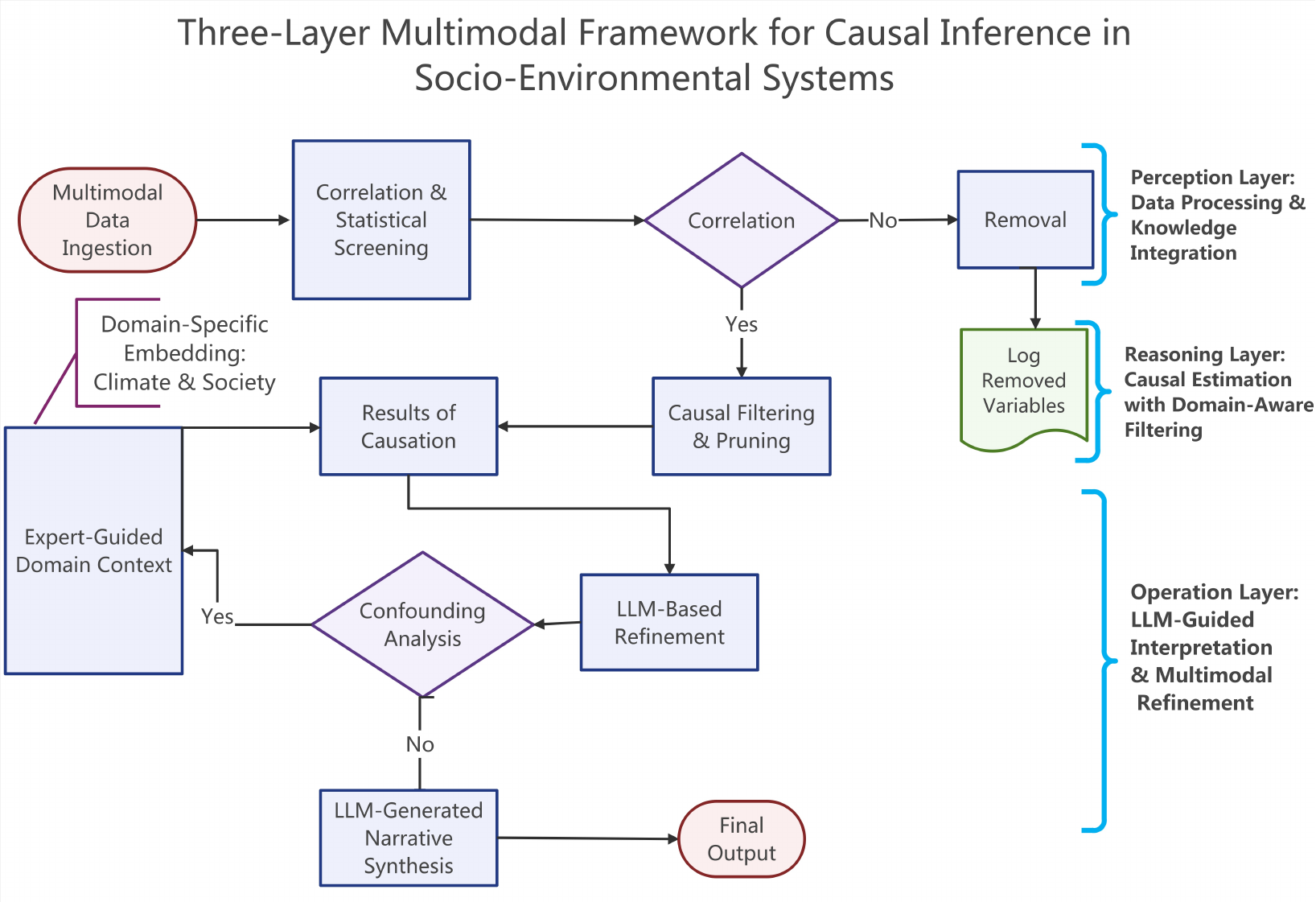}
\caption{Reflexive Multimodal ClimateAgents Framework}
\label{fig:Fig1}
\end{figure*}

ClimateAgents consists of three components: (i) a perception layer that structures multimodal inputs into formal representations (e.g., indicators, semantic frames); (ii) a reasoning layer for planning, inference, and adaptive decision-making; and (iii) an operation layer that performs causal inference, modeling, and policy simulations, with outputs interpreted via LLMs. A continuous agent feedback loop enables real-time refinement and contextual adaptation for evidence-based policy support (Figure~\ref{fig:Fig1}, \ref{fig:Fig3}).

\subsection*{Causal Inference}

Based on former work\cite{shan2024correlation}, this study furhter introduces a \textit{\textbf{three-stage comparative framework}} for investigating causal relationships in the context of social science and climate change, aimed at supporting evidence-based reasoning. The pipeline combines (i) correlation analysis to identify initial statistical associations, (ii) machine learning based causal discovery to estimate structural dependencies, and (iii) LLM-guided prompt exploration to surface contextual explanations and generate policy-relevant hypotheses. Each stage contributes distinct but complementary evidence toward causal interpretation, facilitating more transparent and informed downstream analysis. This approach is designed to support the development of empirical insights that can inform decision-making in complex socio-environmental systems.

\subsection*{Evidence Informed Policy Reasoning}

\begin{figure*}
	\centering
	\includegraphics[width=0.8\linewidth]{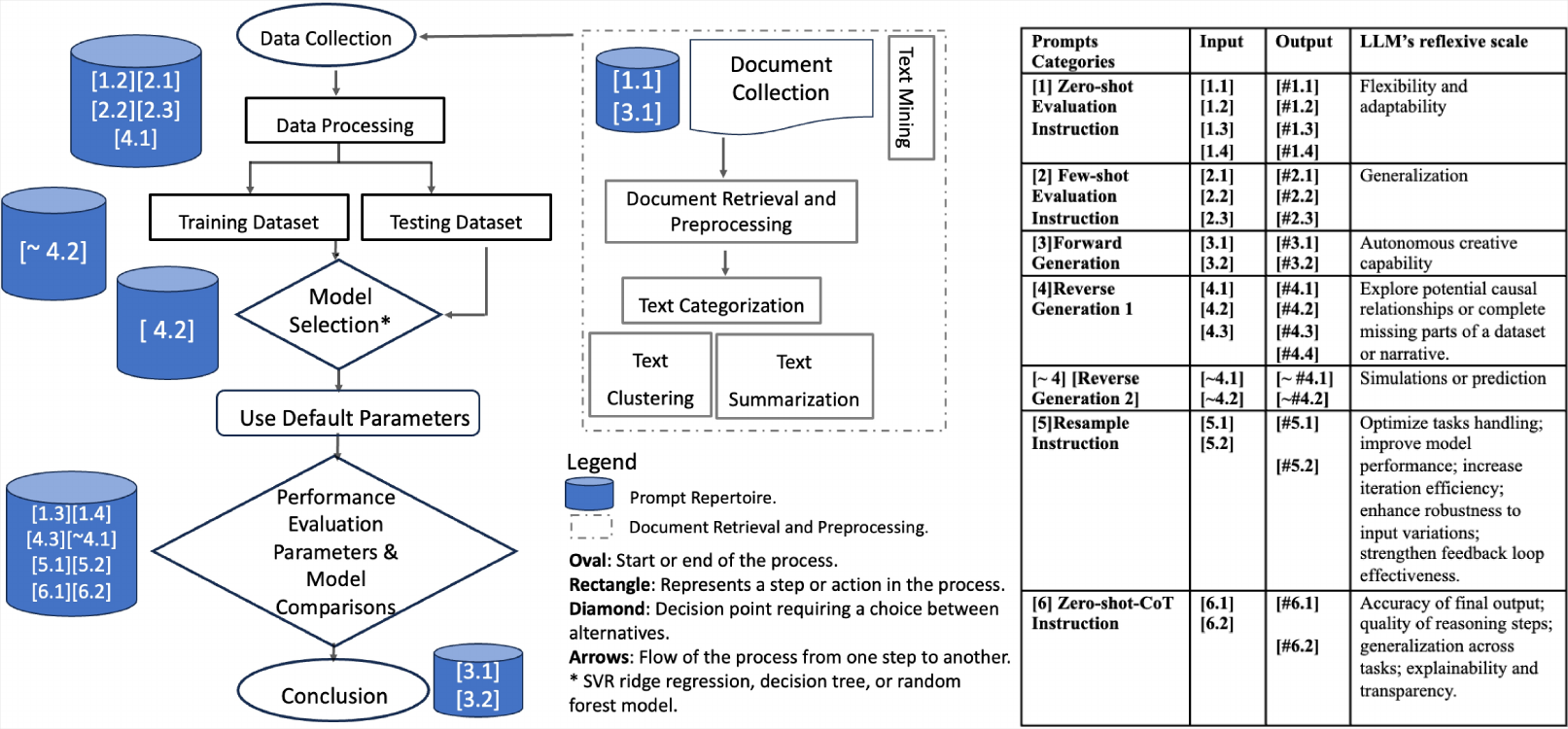}
	\caption{Perception Layer Framework for Climate Change Data Processing and Model Evaluation.}
	\label{fig:Fig2}
\end{figure*}

\textit{\textbf{Evidence retrieval}} is demonstrated through text classification of agent-generated prompts (Figure~\ref{fig:Fig2}), which revealed themes related to carbon emission prediction, including model diversity, geographic specificity, and environmental justice. Using Biopython and the NCBI Entrez database, the system efficiently retrieves and synthesizes relevant literature, supporting large-scale climate and air quality research. \textit{\textbf{Causal effects}} are estimated following Rolland et al. \cite{rolland2022score}, modeling each variable as a function of its causal parents with additive noise. Leaf nodes are identified using score function derivatives, and topological ordering is achieved by sequential leaf removal, with the Jacobian approximated by the  Stein gradient estimator and refined through the CAM procedure \citep{rolland2022score}. \textit{\textbf{Validation and interpretation}} involve domain expertise and standard metrics, with results highlighting key drivers such as rural and urban access to clean fuels and urbanization growth.  To valid  LLMs contribute to causal inference, the framework applies World Bank Development Indicators  and employs a taxonomy of causality \citep{liang2023holisticevaluationlanguagemodels,cui2024odysseycommonsensecausalityfoundational} grounded in Pearl’s Causal Hierarchy \citep{pearl2018why,bareinboim2022pearl}. This taxonomy distinguishes direct, preventative, facilitative, resultative, and influential causal verbs, ensuring LLM inquiries classify relations accurately and avoid confounding \citep{girju2002mining,kozareva2012cause,riaz2014recognizing,nazaruka2020overview,allen2005local,harvey2002getting,wolff2003direct,boas2000resultative,pena2015constructionist,yee1996causal,slovic2007affect,slovic2013risk}. Data on carbon emissions per capita from 265 economies, countries and regions, covering the period from 2000 to 2020, are sourced from Climate Watch Historical GHG Emissions.

\begin{figure*}
	\centering
	\includegraphics[width=0.8\linewidth]{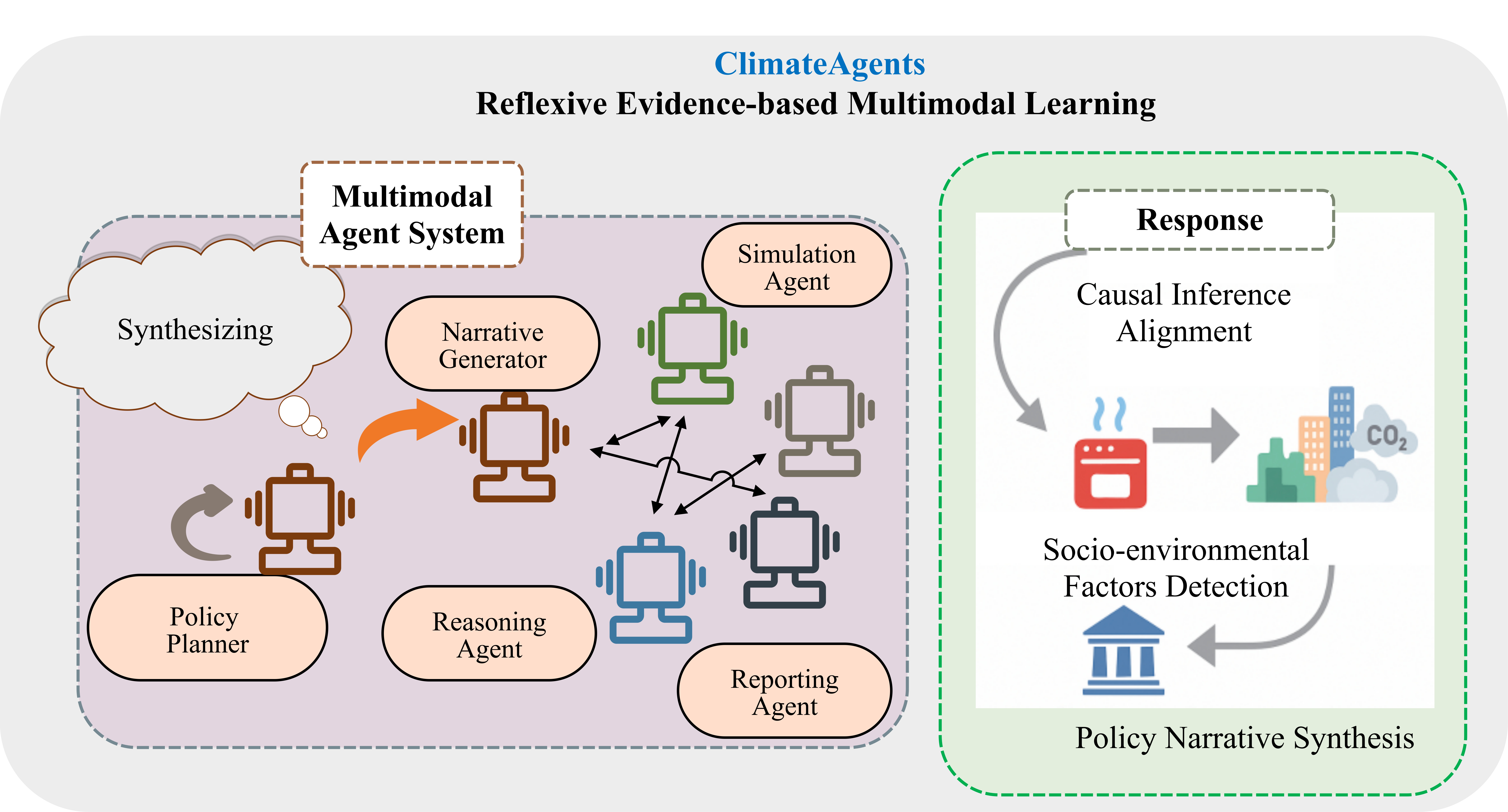}
	\caption{Compact multi-agent framework for policy scenario modeling. Each agent fulfills a specialized role.}
	\label{fig:Fig3}
\end{figure*}

\section*{Results and Discussion}


The multi-agent system initiated the analysis using a planning tool that decomposed complex user queries into actionable subtasks. Execution incorporated multiple tools, including automated statistical computation, literature analysis, knowledge retrieval from scientific databases, and code generation for downstream modeling  (Figure \ref{fig:Fig2}, \ref{fig:Fig3}).

This study identified three key socioeconomic variables with strong causal effects on per capita carbon emissions: (i) \textit{\textbf{Access to clean fuels and technologies for cooking, rural (\% of rural population)}} (EG.CFT.ACCS.RU.ZS), which showed a strong negative relationship with emissions, as higher access reduces biomass reliance and climate impact in less urbanized regions \citep{nathaniel2021environmental, verma2021energy}; (ii) \textit{\textbf{Access to clean fuels and technologies for cooking, urban (\% of urban population)} }(EG.CFT.ACCS.UR.ZS), which similarly lowered emissions in dense populations through improved energy efficiency \citep{naeem2023transitioning}; and (iii) \textit{\textbf{Urban population (\% of total population)}} (SP.URB.TOTL.IN.ZS), which was positively associated with emissions, reflecting increased demand for transport, infrastructure, and industry \citep{hankey2010impacts, madlener2011impacts, li2015impacts}. These results were derived from a causal analysis of 20 years of data across 265 economies, countries and regions, covering 98 socioeconomic indicators. Full steps are provided in SI Appendix.

To evaluate this capacity, the system applied instruction-based prompting techniques, including zero-shot, few-shot, and chain-of-thought (CoT) prompts \citep{Brown2020, zhou2022large, Srivastava2022}. Prompts were anchored to previously identified causal variables—access to clean fuels in rural and urban contexts (EG.CFT.ACCS.RU.ZS and EG.CFT.ACCS.UR.ZS) and urban population percentage (SP.URB.TOTL.IN.ZS)—and structured using a five-part causal taxonomy from the CausalQuest framework \citep{ceraolo2024}, aligned with Pearl’s Causal Hierarchy \citep{pearl2018why, bareinboim2022pearl}. The five causal query types—Direct, Preventative, Facilitative, Resultative, and Influential—were designed to elicit different dimensions of causal reasoning \citep{liang2023holisticevaluationlanguagemodels, cui2024odysseycommonsensecausalityfoundational}.
An illustrative output is shown in Figure~\ref{fig:llm_convo_policy}, where the LLM agent (“AI Social-Scientist”) was prompted to assess whether disparities in clean fuel access might impact the effectiveness of emission-reduction policies. The model responded by highlighting the role of infrastructure inequality, policy generalization limits, and the importance of regional context in designing equitable interventions.

\begin{figure}[H]
	\centering
	\fbox{
		\begin{minipage}{1\textwidth}
			\small
			\textbf{User:} Rural areas (\texttt{EG.CFT.ACCS.RU.ZS}) have lower access to clean fuels than urban ones (\texttt{EG.CFT.ACCS.UR.ZS}), and urbanization (\texttt{SP.URB.TOTL.IN.ZS}) is increasing. 
			Could variability in clean fuel access affect the effectiveness of environmental policies at reducing emissions? \\[0.6em]
			
			\textbf{AI Social-Scientist (GPT-4-turbo):} 
			\textbf{Policy Effectiveness:} Greater variability weakens uniform outcomes.  
			\textbf{Emission Impact:} Urban-focused policies may yield quick results, but lasting success requires rural inclusion.  
			\textbf{Urbanization Factor:} Shifts where and how policies work; less urbanized regions need tailored approaches.  
			\textbf{Policy Design:} Reducing variability can support equitable and impactful strategies.
		\end{minipage}
	}
	\caption{Conversation between a user and an AI social-scientist model (GPT-4-turbo) on clean fuel access and policy effectiveness.}
	\label{fig:llm_convo_policy}
\end{figure}

\section*{Summary and Future Perspectives}

This study introduces \textbf{ClimateAgents}, a multi-agent framework that integrates large language models with a data-driven approach for climate policy analysis. Structured into perception, reasoning, and operation layers, the system combines planning, reasoning, simulation, narrative, and reporting agents to model complex socio-environmental dynamics. By leveraging the reasoning and synthesis capabilities of LLMs alongside computational efficiency, ClimateAgents enables flexible analysis of clean energy, urbanization, and carbon emissions, integrating heterogeneous datasets into interpretable narratives and visualizations that enhance accuracy and support evidence-based policy.

Nonetheless, limitations remain. Causal inference depends on data quality, simplified assumptions about emissions may not generalize, and automated script generation for complex simulations is not yet feasible. Causal inquiry in LLMs is also shaped by social norms and linguistic context \cite{forbes-etal-2020-social,cui2024odysseycommonsensecausalityfoundational}. Future integration of advanced foundation models and generative tools could enhance scalability, reduce computational costs, and expand predictive depth, establishing ClimateAgents as a flexible platform for intelligent climate policy simulation.

\section {Ackowledgement}
The author gratefully acknowledges Prof. Fei Wu for his guidance on the causal framework, Dr. Anpeng Wu for his support with causal analysis, and Prof. Hao Chen for his direction in developing the multimodal framework.

\bibliographystyle{unsrt}
\bibliography{pnas-sample.bib}

\appendix

\section*{Supplementary Information}


%

\label{sec:appendix}

%

\subsection*{Knowledge Retrieval} Text classification of agent-generated prompts ([1.1], [3.1] of Figure \ref{fig:Fig2} revealed key themes related to carbon emission prediction. The extracted summaries highlighted model diversity, geographic specificity, environmental justice considerations, and challenges in temporal and spatial scaling. The study utilizes the Biopython library and the NCBI Entrez database to provide an example of literature retrieval and analysis. Running this code snippet allows for efficient retrieval of article details, including titles, from the PubMed database, facilitating in-depth analysis and synthesis of research findings. It offers a tool for researchers investigating the complex relationship between climate change and global air quality, enabling the retrieval and analysis of a large volume of articles. These results demonstrate the system's ability to identify cross-cutting concerns in climate-related modeling efforts.

\subsection* {Causal Effects Estimation}

The method employed in this research is adapted from existing approaches to causal modeling, specifically following the framework outlined by Rolland et al. \cite{rolland2022score}. In this approach, each variable is modeled as a function of its direct causal parents in the causal graph, along with an additive noise term. The data distribution is defined by these causal relationships, and score functions are used to identify leaf nodes within the graph. Leaf nodes are detected based on the variance of partial derivatives of the score function, which helps distinguish parent-child relationships among variables. The nodes in the graph are arranged in order by finding and removing leaf nodes one by one. The Jacobian of the score can be approximated using the Stein gradient estimator with ridge RBF kernel regression \cite{rolland2022score}.

\subsection*{Causal Graph Construction and Score Matching}

Based on this finding provided by Rolland et al., 2022, the experiment achieves topological ordering by sequentially identifying the leaf nodes and removing them one by one. The Jacobian of the score can be approximated by Stein gradient estimator with ridge RBF kernel regression \citep{rolland2022score}.

Once a topological order is estimated, the causal graph is constrained to be a subgraph of a fully connected graph. However, pruning is necessary to remove spurious edges, which is achieved using the CAM pruning process.

\subsection*{CAM Pruning}

The methods described above control for confounding variables by retaining key confounders during variable selection and removing irrelevant variables through correlation analysis. The CAM pruning process refines the causal graph by eliminating spurious relationships while preserving causal integrity.

After arranging the nodes, the graph is refined by using the CAM pruning process, which removes unnecessary connections to reveal the actual causal structure, aligning with methods discussed by Rolland et al.\citep{rolland2022score}. Detailed outputs are included the following metrics \footnote{It is noted that "Variable Selection" is to ensure that important confounders are included before pruning begins, as removing key variables early can lead to residual confounding or spurious relationships. The formal analysis of correlation removes unrelated variables-those that have no meaningful relationship with the target variable or the other variables in the system. These variables are unlikely to act as confounders since they do not introduce residual confounding or spurious relationships when removed.
	
	For validation, after CAM pruning, the causal structure is validated using domain expertise to ensure the robustness of the inferred causal graph. CAM pruning is not a substitute for confounding control methods. It is suggested to be used in combination with other techniques to ensure the validity of causal inferences. This is also the rationale for incorporating LLMs with expertise knowledge for further exploration.}:

\begin{itemize}
	\item Structural similarity: Evaluated using SID and SHD.
	\item Predictive accuracy: Measured through precision, recall, and F1 score.
	\item Overall deviation: Assessed using L2 distance.
\end{itemize}

 By highlighting key variables—such as access to clean fuels in rural and urban areas and urban population growth—the graph supports more targeted and effective policy-making.

\section* {Validation: From Correlations to Causation via LLM Inquiries}

In the specific context of climate change, do LLMs offer better causal inference? To address the request involving the exploration of causality factors for carbon emissions using the World Bank variables ("EG.CFT.ACCS.RU.ZS", "EG.CFT.ACCS.UR.ZS", "SP.URB.TOTL.IN.ZS") as the prior benchmark, the study categorizes questions into five main types for LLMs causality taxonomy prompts described in Txonomy of Causality. The study follows \citep{ceraolo2024}'s CausalQuest database, with a focus on the economic and climate change context. Similarly, the study adopts Pearl’s Causal Hierarchy (PCH) framework (\citep{pearl2018why,bareinboim2022pearl}), and defines a causal question as one that meets the following criteria: a question is considered causal if it involves, or if its solution process includes, any inquiry into the effects given a specific cause, and the causes given a specific effect, or the causal relationship between the given variables.

\subsection *{Taxonomy of Causality}
The causal taxonomy-"Direct, Preventative, Facilitative, Resultative, and Influential"-describes various types of causal relationships that verbs can imply (see SI-Table \ref{tab:causal_taxonomy} ). This approach controls for confounding variables during LLM inquiries by leveraging a structured causal taxonomy to identify, classify, and account for different types of causal relationships\citep{liang2023holisticevaluationlanguagemodels,cui2024odysseycommonsensecausalityfoundational}.
\begin{table}[ht]
	\caption{Causal taxonomy used in LLM-based inquiries}
	\label{tab:causal_taxonomy}
	\centering
	\small
	\begin{tabular}{p{2.5cm} p{9.5cm}}
		\toprule
		\textbf{Category} & \textbf{Description and Example} \\
		\midrule
		\textbf{Direct} & Refers to actions or driving forces that have an immediate impact on outcomes. The cause directly influences the effect without intermediaries. Typical verbs: ``increase'', ``trigger'' \citep{girju2002mining,kozareva2012cause,riaz2014recognizing,nazaruka2020overview}. \newline Example: Urban access to clean fuels directly reduces carbon emissions. \\
		
		\textbf{Preventative} & Causes that reduce or prevent the likelihood of a negative outcome. Verbs: ``prevent'', ``reduce'', ``inhibit'' \citep{allen2005local}. \newline Example: Improved access to clean technologies prevents an increase in carbon emissions. \\
		
		\textbf{Facilitative} & Causes that enable or support an effect without directly causing it. Verbs: ``enable'', ``allow'', ``support'' \citep{harvey2002getting,wolff2003direct}. \newline Example: Access to urban clean fuels facilitates a reduction in carbon emissions. \\
		
		\textbf{Resultative} & Causes that lead to specific outcomes, emphasizing consequences. Verbs: ``lead to'', ``result in'', ``cause'' \citep{boas2000resultative,pena2015constructionist}. \newline Example: Urban population increase results in higher carbon emissions. \\
		
		\textbf{Influential} & Factors that indirectly affect the likelihood or intensity of an outcome. Verbs: ``influence'', ``impact'', ``affect'' \citep{yee1996causal,slovic2007affect,slovic2013risk}. \newline Example: Urbanization influences carbon emissions through changes in energy use patterns. \\
		\bottomrule
	\end{tabular}
\end{table}

\end{document}